\documentclass[cameraready]{Interspeech}

\title{Cloud-Boosted Low-Compute Multi-Channel Speech Enhancement}

\author[affiliation={1\dagger},internship]{Xulin}{Fan}
\author[affiliation={2}]{Juan}{Azcarreta}
\author[affiliation={2}]{Ashutosh}{Pandey}
\author[affiliation={2}]{Jesus}{Alvarez}
\renewcommand{\authorsep}{,\\}
\author[affiliation={2}]{Ke}{Tan}
\author[affiliation={2}]{Jacob}{Donley}
\author[affiliation={2}]{Ritwik}{Giri}
\author[affiliation={2}]{Buye}{Xu}

\address{
    $^1$ University of Illinois Urbana-Champaign, USA \\
    $^2$ Meta Reality Labs Research, USA 
}

\email{xulinf2@illinois.edu, jazcarretao@meta.com, apandey620@meta.com, jsmalvarez@meta.com, tanke1116@meta.com, jdonley@meta.com, ritwikgiri@meta.com, xub@meta.com}


\usepackage{comment}
\usepackage{booktabs}
\usepackage{multirow}
\usepackage{makecell}
\usepackage{graphicx} 
\usepackage[table]{xcolor} 

\begin{document}

\maketitle

\begin{abstract}
Low-latency, low-compute speech enhancement is essential for wearable devices with real-time communication requirements, but strict computational constraints significantly limit on-device performance. Knowledge Boosting has been proposed as an effective approach to improve edge model performance by leveraging a more capable server-side model, but performance gains for speech enhancement have been limited. We propose a collaborative framework incorporating three techniques: (1) delayed server output as additional input, (2) layerwise feature boosting that transfers intermediate server representations to guide edge inference, and (3) collaborative multichannel Wiener filtering, which fuses weighted covariance matrices estimated from both server and edge models for improved beamforming. Experimental results demonstrate that the proposed collaborative framework significantly outperforms the edge-only baseline with minimal additional computational overhead.

\end{abstract}

\section{Introduction}

The ubiquitous deployment of wearable devices, such as smart glasses and hearables, has catalyzed a growing demand for high-fidelity real-time audio processing. Among the myriad of audio tasks, on-device speech enhancement is particularly critical for enabling clear communication in noisy and reverberant environments. To meet strict latency thresholds and hardware constraints (e.g., battery life, thermal limits), the research community has developed a variety of compact speech enhancement models~\cite{westhausen2023low, patel2023uxnet, schroter2022deepfilternet2, pandey2025tinygru, cheng2025slowfast}. While these lightweight architectures achieve impressive efficiency, their reduced parameter counts and shallow depths often fundamentally limit their capacity to model complex acoustic scenes, resulting in performance that lags significantly behind server-level solutions.

Conversely, the state-of-the-art in speech enhancement is dominated by high-capacity deep neural networks~\cite{quan2024spatialnet, wang2023tfgridnet, zhang2024improving, zhang2023use}. By leveraging large-scale spatial-spectral attention mechanisms and massive parameter sets, these models demonstrate remarkable robustness even under ultra-low signal-to-noise ratio (SNR) conditions. However, the excessive computational cost and memory footprint of such models render them prohibitive for deployment on edge devices, creating a distinct performance gap between cloud-based and on-device processing.

To bridge this gap, hybrid approaches~\cite{wang2022neuralbeam1, luo2022hybrid, kuang2023hybrid2, pandey2025tinygru, hsieh2024nwf, lee2023neuralbeam2, wang2020dereverb} have gained popularity. These methods integrate traditional signal processing, specifically spatial filtering, with neural networks. By offloading spatial discrimination to mathematical beamformers (e.g., minimum-variance distortionless response (MVDR)~\cite{doclo2010mvdr} or multichannel Wiener filter (MCWF)~\cite{wang2021mcwf}) and using small neural networks primarily for mask estimation, these systems benefit from the stability of physical acoustic models. Nevertheless, the performance of hybrid systems remains bounded by the accuracy of the statistics estimated by the lightweight neural estimator. If the edge model fails to identify the target in a complex scene, the resulting beamformer collapses.


Recently, Knowledge Boosting~\cite{srinivas2024knowledgeboost} was introduced as a paradigm to enhance edge model performance by leveraging a powerful server-side model via a network connection. Assuming a fixed communication delay, this framework trains the edge and server models collaboratively. While this approach has yielded significant gains for target speaker extraction and separation, improvements for general speech enhancement have remained modest. This suggests that simply conditioning an edge model on server outputs is insufficient to capture the rapid spectral variations of non-stationary noise when communication latency is present. However, two critical questions remain unexplored: (1) what types of information from the server are most beneficial for improving edge model performance under communication delays and minimal edge-side computation, and (2) how knowledge boosting can be integrated into hybrid beamforming approaches, where server-side information assists the edge device in computing a robust spatial filter.

In this paper, we propose a comprehensive collaborative framework designed to maximize the utility of a frozen, pretrained server-side model for improving edge-side inference. Unlike~\cite{srinivas2024knowledgeboost}, which trains both models jointly and relies solely on delayed output conditioning, our framework keeps the server fixed and introduces intermediate representation-level guidance and spatial-statistics fusion. Our core insight is that while spectral details may change rapidly, spatial statistics (such as the target speaker's direction of arrival) evolve much more slowly and are robust to communication delays. We introduce three complementary techniques to exploit this: (1) \textit{Delayed Input Concatenation}, where the server's enhanced audio is fed as an auxiliary reference, providing a clean but delayed prior to the edge model; (2) \textit{Layerwise Feature Boosting}, which transfers hierarchical representations via Feature-wise Linear Modulation (FiLM)~\cite{perez2018film} to guide the edge model's internal states; and (3) \textit{Collaborative Multichannel Wiener Filter (MCWF)}. We propose a novel fusion strategy where covariance matrices from both the server and edge are combined. This allows the system to leverage the superior spatial selectivity of the server model to stabilize the on-device beamformer.

We evaluate this framework under challenging low-SNR conditions and a more than 64ms server-edge communication delay. Experimental results demonstrate that our collaborative approach significantly outperforms strong edge-only baselines, closing the gap with server-grade performance while incurring negligible ($<$5\%) computational overhead on the edge device.

\section{Method}

\subsection{Overview}

Our proposed collaborative speech enhancement system consists of a high-capacity server-side model and a lightweight edge-side model. The server-side model is first pretrained on the speech enhancement task and subsequently frozen during edge model training and inference. During operation, the server-side model processes the multichannel input and produces auxiliary outputs, including enhanced spectrograms and intermediate layer representations, that are transmitted to the edge device to boost the performance of the edge-side model.

\begin{figure}[t]
    \centering
    \includegraphics[width=\linewidth]{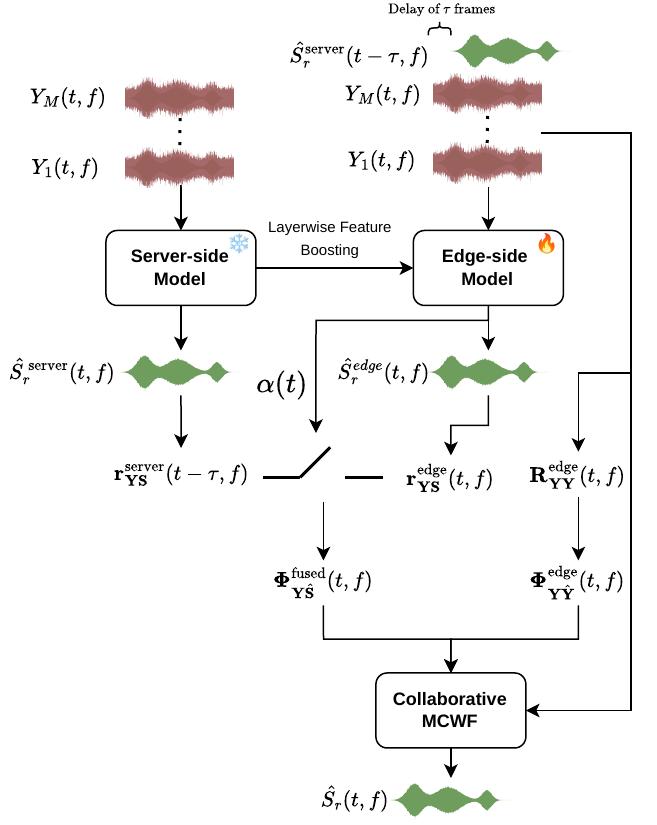}
    \caption{Overview of the proposed collaborative speech enhancement system. The server-side model processes multichannel input and transmits enhanced spectrograms and intermediate representations to the edge-side model via a communication channel.}
    \label{fig:system_overview}
    \vspace{-0.4cm}
\end{figure}

\subsection{Server-side Model}

The server-side model is a causal SpatialNet~\cite{quan2024spatialnet}, a deep neural network designed for multichannel speech enhancement. The architecture processes time-frequency representations of the microphone array input through a stack of transformer-based narrow-band and cross-band modules. The network maintains causality by masking future frames in all attention and convolution operations, ensuring that each output frame depends only on current and past inputs. The network is pretrained on the speech enhancement task and remains frozen during training and inference of the edge-side model.

We identify two types of auxiliary information from SpatialNet that can improve edge-side performance. The first is the final enhanced spectrogram, representing SpatialNet's estimate of the target speech. The second is a set of intermediate layer representations extracted at selected depths within the network. These intermediate features capture hierarchical spatial and spectral patterns, from low-level acoustic cues in early layers to high-level source characteristics in deeper layers, providing multi-scale guidance to the edge model.

\subsection{Edge-side Model}

The edge-side model is based on TinyGRU~\cite{pandey2025tinygru}, a compact architecture designed for efficient real-time inference on resource-constrained hardware. The network consists of a spatial convolution module that reduces the multichannel input to a single-channel representation, followed by a temporal processing module comprising three stacked SplitGRU layers. The network outputs a complex-valued mask that is applied to the noisy input spectrogram to obtain an initial clean speech estimate.

This masked estimate is then used to derive a multichannel Wiener filter (MCWF) beamformer, which produces the final enhanced output by optimally combining information across all microphone channels. The MCWF formulation enables the lightweight edge model to leverage spatial diversity in the microphone array, achieving superior enhancement compared to single-channel masking alone.

To incorporate server-side guidance at the earliest processing stage, we concatenate the delayed enhanced spectrogram from SpatialNet to the multichannel input before the spatial convolution module. Specifically, the server-side enhanced output $\hat{S}_{\text{server}}(t-\tau, f)$, delayed by $\tau$ frames to account for communication latency, is appended as additional input channels alongside the $M$-channel microphone input. This early conditioning provides the edge model with a direct reference signal that captures the server's estimate of the target speech, allowing the spatial convolution to leverage this information when learning to reduce the multichannel input to a single-channel representation. Since we stack real and imaginary parts, the M-channel microphone input yields 2M real-valued channels, and the complex-valued server output contributes 2 additional channels, resulting in 2(M+1) input channels total.

\subsection{Layerwise Feature Boosting}

To enable knowledge transfer from the server-side SpatialNet to the device-side TinyGRU, we extract intermediate representations from multiple depths of the SpatialNet rather than relying solely on its final output. Specifically, we sample features from $N=4$ points: the input encoding layer (before any transformer processing), and SpatialNet layers 4, 8, and 12. Let $\mathbf{h}^{(l)} \in \mathbb{R}^{B \times F \times T \times H}$ denote the feature extracted from layer $l$, where $B$ is the batch size, $F$ the number of frequency bins, $T$ the number of time frames, and $H$ the hidden dimension. Each feature is compressed via a learned $1 \times 1$ convolution that projects the hidden dimension to a scalar, yielding $\mathbf{c}^{(l)} \in \mathbb{R}^{B \times T \times F}$. These compressed features are temporally delayed by $\delta$ frames to account for server-side processing latency.

The processed features are injected into the $N$-layer TinyGRU through hierarchical Feature-wise Linear Modulation (FiLM)~\cite{perez2018film}. The input encoder features modulate the input to GRU layer 1, layer 4 features modulate the input to GRU layer 2, layer 8 features modulate the input to GRU layer 3, and layer 12 features modulate the final GRU output. Let $\tilde{\mathbf{c}}_j \in \mathbb{R}^{B \times T \times F}$ denote the $j$-th conditioning feature after compression and delay. At each conditioning point $j \in \{1, \ldots, N\}$, a linear projection $\mathbf{W}_j \in \mathbb{R}^{2d \times F}$ generates scale and shift parameters independently for each time frame $t$:

\begin{equation}
[\boldsymbol{\gamma}_j(t), \boldsymbol{\beta}_j(t)] = \mathbf{W}_j \tilde{\mathbf{c}}_j(t) + \mathbf{b}_j
\end{equation}
where $d$ is the RNN hidden dimension. The modulated representation is:

\begin{equation}
\tilde{\mathbf{x}}_j(t) = \sigma(\boldsymbol{\gamma}_j(t)) \odot \mathbf{x}_j(t) + \boldsymbol{\beta}_j(t)
\end{equation}
where $\sigma(\cdot)$ is the sigmoid function and $\odot$ denotes element-wise multiplication. This hierarchical design allows early SpatialNet features to provide low-level spectral conditioning, while deeper features capture increasingly refined speech-noise separation cues at later processing stages.

\begin{table*}[t]
\centering
\caption{Performance comparison of edge-side speech enhancement models on standard and challenging conditions. We incrementally add three improvements to the baseline: (a) delayed server output conditioning, (b) layerwise feature boosting, and (c) collaborative MCWF. SI-SDR is in dB; STOI is in \%. Parameters (in thousands) and the number of multiply-accumulate operations in millions (MMACs) are reported for edge-side only. Causal SpatialNet (in gray) serves as a high-capacity server-side reference. Best  results under 64ms server-edge delay are in \textbf{bold}.}
\vspace{-0.3cm}
\label{tab:results}
\resizebox{\linewidth}{!}{%
\begin{tabular}{l|c|cc|ccc|ccc}
\toprule
\multirow{2}{*}{Model} & \multirow{2}{*}{\makecell{Server-Edge\\Delay (ms)}} & \multirow{2}{*}{Params (K)} & \multirow{2}{*}{MMACs} & \multicolumn{3}{c|}{Standard} & \multicolumn{3}{c}{Challenging}\\
\cmidrule(lr){5-7} \cmidrule(lr){8-10}
 & & & & SI-SDR $\uparrow$ & PESQ $\uparrow$ & STOI $\uparrow$ & SI-SDR $\uparrow$ & PESQ $\uparrow$ & STOI $\uparrow$ \\
\midrule
Noisy Input & -- & -- & -- & -4.96 & 1.68 & 69.27 & -10.01 & 1.28 & 55.70 \\
\midrule
TinyGRU+MCWF~\cite{pandey2025tinygru} & -- & 145.1 & 32.9 & 1.97 & 2.29 & 82.36 & -1.16 & 1.90 & 70.49 \\
+(a) (proposed)& 64 & 147.2 (+1.4\%) & 33.2 (+0.9\%) & 3.05 & 2.34 & 83.93 & -0.02 & 1.97 & 72.58 \\
+(a,b) (proposed)& 64 & 147.2 (+1.4\%) & 33.3 (+1.2\%) & 4.47 & \textbf{2.35} & 84.24 & 1.35 & \textbf{1.98} & 72.68 \\
+(a,b,c) (proposed)& 64 & 147.3 (+1.5\%) & 33.7 (+2.4\%) & \textbf{5.74} & 2.33 & \textbf{85.48} & \textbf{2.33} & 1.95 & \textbf{73.73} \\
TinyGRU-Large+MCWF~\cite{pandey2025tinygru} & -- & 432.4 (+198.0\%) & 68.77 (+109.0\%) & 2.75 & 2.33 & 83.54 & -0.37 & 1.95 & 72.19 \\
\midrule
+(a,b,c) (proposed)& 96 & 147.3 (+1.5\%) & 33.7 (+2.4\%) & 4.39 & 2.34 & 84.16 & 1.14 & 1.96 & 72.31 \\
+(a,b,c) (proposed)& 128 & 147.3 (+1.5\%) & 33.7 (+2.4\%) & 4.26 & 2.34 & 84.13 & 1.07 & 1.96 & 72.22 \\
\midrule
\textcolor{gray}{Causal SpatialNet~\cite{quan2024spatialnet}} & -- & \textcolor{gray}{--} & \textcolor{gray}{--} & \textcolor{gray}{11.79} & \textcolor{gray}{3.35} & \textcolor{gray}{95.47} & \textcolor{gray}{9.08} & \textcolor{gray}{2.97} & \textcolor{gray}{91.53} \\
\bottomrule
\end{tabular}%
}
\vspace{-0.5cm}
\end{table*}

\subsection{Collaborative MCWF}

In the edge-only approach, the device-side TinyGRU model estimates a target speech signal, which is used to compute covariance statistics for deriving the MCWF beamformer coefficients. However, the lightweight edge model may produce suboptimal target estimates in challenging acoustic conditions. To address this limitation, we propose a collaborative MCWF framework that leverages target estimates from both the device-side TinyGRU and the server-side SpatialNet, fusing their statistical contributions to compute more robust beamformer coefficients.

\subsubsection{Frame-wise Outer Product Computation}

At each time-frequency bin, we compute the instantaneous spatial covariance matrix from the multichannel input:
\begin{equation}
\mathbf{R}_{\mathbf{YY}}(t,f) = \mathbf{Y}(t,f) \mathbf{Y}^H(t,f)
\end{equation}
where $\mathbf{Y}(t,f) \in \mathbb{C}^{M}$ is the $M$-channel input STFT at time frame $t$ and frequency bin $f$. Since this depends only on the input signal (identical on edge and server), it is computed solely on the edge.

The cross-covariance vector between the input and estimated target is computed independently by both models using their respective target estimates $\hat{\textbf{S}}(t,f)$:
\begin{equation}
\mathbf{r}_{\mathbf{YS}}(t,f) = \mathbf{Y}(t,f) \hat{\textbf{S}}^{H}(t,f)
\end{equation}
yielding $\mathbf{r}_{\mathbf{YS}}^{\text{edge}}(t,f)$ and $\mathbf{r}_{\mathbf{YS}}^{\text{server}}(t,f)$ for the edge and server models, respectively.

\subsubsection{Cross-Covariance Fusion}

Due to communication latency, the server-side statistics arrive with a delay of $\tau$ frames. This creates a fundamental trade-off: the server-side SpatialNet produces higher-quality target estimates due to its greater model capacity, but these estimates reflect the acoustic scene from $\tau$ frames in the past. Conversely, the edge-side TinyGRU produces lower-quality estimates but captures the current acoustic conditions without delay. We fuse the cross-covariance vectors using an adaptive weight to navigate this trade-off:
\begin{equation}
\mathbf{r}_{\mathbf{YS}}^{\text{fused}}(t,f) = (1 - \alpha(t)) \cdot \mathbf{r}_{\mathbf{YS}}^{\text{server}}(t-\tau,f) + \alpha(t) \cdot \mathbf{r}_{\mathbf{YS}}^{\text{edge}}(t,f)
\end{equation}

%
where $\alpha(t) \in [0,1]$ is a per-frame scalar predicted by TinyGRU's final linear layer and applied uniformly across all frequency bins. This weight controls the balance between server and edge contributions: when $\alpha$ approaches 0, the system favors the more accurate but delayed server-side statistics; when $\alpha$ approaches 1, it favors the less accurate but current edge-side statistics.

The network learns to adapt this weight based on acoustic conditions. In stationary or slowly-varying environments, where the acoustic scene changes little over $\tau$ frames, the delayed server statistics remain highly relevant, and lower $\alpha$ values allow the system to benefit from SpatialNet's superior estimation quality. In contrast, during rapid acoustic changes such as sudden noise onsets or abrupt spatial transitions, the delayed server statistics may no longer reflect the current scene, and higher $\alpha$ values allow the system to rely on the edge model's real-time estimates despite their lower quality.

\subsubsection{Time-Varying Smoothing}

The frame-wise statistics are accumulated over time through learned recursive smoothing. Dropping the time-frequency indices $(t,f)$ for brevity, the spatial covariance matrix and fused cross-covariance vector are updated as: \begin{equation} \boldsymbol{\Phi}_{\mathbf{YY}} = (1 - \beta_{\mathbf{YY}}) \cdot \boldsymbol{\Phi}_{\mathbf{YY}}^{\text{prev}} + \beta_{\mathbf{YY}} \cdot \mathbf{R}_{\mathbf{YY}} \end{equation} \begin{equation} \boldsymbol{\Phi}_{\mathbf{YS}} = (1 - \beta_{\mathbf{YS}}) \cdot \boldsymbol{\Phi}_{\mathbf{YS}}^{\text{prev}} + \beta_{\mathbf{YS}} \cdot \mathbf{r}_{\mathbf{YS}}^{\text{fused}} \end{equation} where $\boldsymbol{\Phi}^{\text{prev}}$ denotes the value at the previous frame $(t-1,f)$.

%
The smoothing coefficient $\beta(t,f)$ is computed as the product of a per-frame scalar weight $w(t)$ predicted by the TinyGRU and a learned per-frequency weight vector $\mathbf{v} \in \mathbb{R}^F$ which add temporal variability upon per-frequency smoothing~\cite{grinstein2025smoothing} :
\begin{equation}
\beta(t,f) = \sigma\left( \mathbf{v} \cdot \text{softplus}(w(t)) \right)
\end{equation}
where $\sigma(\cdot)$ is the sigmoid function. This factorization allows the network to learn frequency-dependent smoothing characteristics (e.g., longer integration for low frequencies) while the per-frame scalar enables adaptation to temporal dynamics. Separate weight pairs $(w_{\mathbf{YY}}, \mathbf{v}_{\mathbf{YY}})$ and $(w_{\mathbf{YS}}, \mathbf{v}_{\mathbf{YS}})$ are used for the spatial covariance and cross-covariance, respectively.

\subsubsection{Beamformer Computation}

The MCWF coefficients are computed from the smoothed statistics:
\begin{equation}
\mathbf{W}(t,f) = \mathbf{\Phi}_{\mathbf{YY}}^{-1}(t,f) \cdot \mathbf{\Phi}_{\mathbf{YS}}(t,f)
\end{equation}
and the enhanced signal is calculated as:
\begin{equation}
\hat{\textbf{S}}_{\text{MCWF}}(t,f) = \mathbf{W}^H(t,f) \mathbf{Y}(t,f)
\end{equation}

\section{Experiments}

\subsection{Data}\label{sec:data}

We use clean speech and noise sources from the DNS-Challenge dataset~\cite{reddy2020interspeech}. An 8-channel circular microphone array is employed to simulate multichannel audio input, with room impulse responses (RIRs) generated using Pyroomacoustics. For each sample, we randomly sample room dimensions ranging from $3\times3\times2$ to $10\times10\times5$ meters, with the microphone array placed at a random position within the room. The wall absorption coefficient is uniformly sampled from $[0.1, 0.7]$. With a probability of 0.5, we include 8--16 interference speakers to simulate babble noise. The number of diffuse noise sources is sampled from $[1, 10]$. The signal-to-interference ratio (SIR) is sampled from $[5, 10]$~dB.

We create two datasets of varying difficulty, differing only in signal-to-noise ratio (SNR): the \textit{Standard} dataset uses SNR sampled from $[-5, 10]$~dB, while the \textit{Challenging} dataset uses SNR sampled from $[-10, -5]$~dB. For each dataset, we simulate 80,000 samples for training, 2,000 for validation, and 4,000 for testing.

\subsection{Experimental Setting}

We select a vanilla TinyGRU+MCWF~\cite{pandey2025tinygru} with time-varying smoothing as the baseline. For experiments with server-side boosting, we first pretrain the server-side SpatialNet model for 100 epochs with PCM loss \cite{pandey2022PCM} until convergence. The server-side model is then frozen and the edge-side TinyGRU model is trained with the respective boosting strategy.

Both the baseline and TinyGRU with server-side boosting are trained for 100 epochs using the Adam optimizer with AMSGrad, an initial learning rate of $10^{-3}$, and gradient clipping with a maximum norm of 0.03. The learning rate is decayed by a factor of 0.1 at epochs 50 and 80 following a multi-step schedule. The training objective is the signal-to-noise ratio (SNR) loss computed between the enhanced output and the clean target signal. Both training and evaluation use the anechoic target signal as the reference.

The audio is processed at 16kHz with a short-time Fourier transform using a 256-sample (16ms) frame size and 128 sample (8ms) hop size, resulting in 129 frequency bins. Training samples are 4-second segments randomly chunked from longer recordings, with a batch size of 32. To simulate the round-trip communication latency between the edge device and server, a delay of $\tau$ frames ($8\tau$ms) is applied to all server-side outputs/features before they are transmitted to the edge model.

\subsection{Results}

Table~\ref{tab:results} presents performance comparisons on Standard and Challenging sets. We analyze the contribution of each proposed component and examine the impact of server-edge communication latency.

\textbf{Baseline Performance.}
The baseline TinyGRU+MCWF model achieves reasonable enhancement on the Standard dataset, improving SI-SDR from -4.96~dB to 1.97~dB and STOI from 69.27\% to 82.36\%. However, performance degrades substantially under the Challenging condition (lower SNR), where SI-SDR reaches only -1.16~dB when the noisy input is -10.01~dB. This highlights the limitations of edge-only models in adverse acoustic conditions.

\textbf{Ablation Study.}
We incrementally evaluate each proposed component. Adding delayed server output conditioning~(a) improves SI-SDR from 1.97~dB to 3.05~dB on Standard and from -1.16~dB to -0.02~dB on Challenging, confirming that delayed server-side clean estimate benefits the edge model. Incorporating layerwise feature boosting~(b) further improves SI-SDR to 4.47~dB and 1.35~dB on Standard and Challenging, respectively, demonstrating that multi-layer feature conditioning helps the edge model better leverage server information at different depth of the server model. The full model with collaborative MCWF~(c) achieves the best performance: 5.74~dB SI-SDR on Standard and 2.33~dB on Challenging, representing total improvements of 3.77~dB and 3.49~dB over baseline.

\textbf{Comparison with Scaled-Up Baseline.}
To verify that improvements stem from the server-edge architecture rather than increased model capacity, we compare against TinyGRU-Large, where we simply scale up the hidden dimension of TinyGRU from 96 to 192, which results in 198\% more parameters and 109\% more MMACs. Despite this substantial increase, TinyGRU-Large achieves only 2.75~dB and -0.37~dB SI-SDR on Standard and Challenging, which significantly underperforms our proposed boosting method (5.74~dB and 2.33~dB) which adds only 1.5\% parameters. This demonstrates that server-side information provides benefits that cannot be replicated by scaling edge computation alone.

\textbf{Server-Edge Delay Sensitivity.}
At 64~ms server-edge delay, the model achieves optimal performance. Increasing latency to 96~ms and 128~ms results in moderate degradation (SI-SDR drops from 5.74~dB to 4.39~dB and 4.26~dB on Standard), but performance remains substantially above the baseline. This graceful degradation indicates the model can effectively leverage delayed server guidance under realistic network conditions.


\section{Conclusion}

We presented a collaborative speech enhancement framework that bridges the gap between lightweight edge models and high-capacity server models by exploiting delay-tolerant spatial statistics. Our approach integrates delayed server output conditioning, layerwise feature boosting, and collaborative MCWF to transfer spatial knowledge from a pretrained server-side SpatialNet to a compact edge-side TinyGRU. Experiments demonstrate SI-SDR improvements of 3.77~dB and 3.49~dB on standard and challenging conditions, respectively, while adding only 1.5\% parameters and 2.4\% computation on the edge. These results show that server-edge collaboration offers a promising path toward high-quality real-time speech enhancement on resource-constrained devices.

\section{Generative AI Use Disclosure}
We used GPT and Claude for language editing and manuscript polishing, including improving clarity, correcting grammatical errors, and formatting \LaTeX{} tables. All technical content, experimental design, analysis, and scientific contributions are entirely the work of the authors. The authors have carefully reviewed the manuscript and take full responsibility for its content.
\bibliographystyle{IEEEtran}
\bibliography{mybib}

\end{document}